\documentclass[aps,prl,twocolumn,floats,showpacs,floatfix,a4paper]{revtex4}
\usepackage{epsfig}
\usepackage{color}
\usepackage{bm}
\usepackage{latexsym}
\usepackage{amssymb}

\begin{document}

\bibliographystyle{unsrt}

\newcommand{\half}{\mbox{\small $\frac{1}{2}$}}
\newcommand{\bra}[1]{\left\langle #1\right|}
\newcommand{\ket}[1]{\left|#1\right\rangle }
\newcommand{\acos}[1]{\mbox{a}\cos }

%%%%%%%%%%%%%%%%%%%%%%%%%%%%%%%%%%%%%%%%%%%%%%%%%%%%%%%%%%%%%%%%%%%%%%%%
\title{One Parameter Scaling Theory for Stationary States of Disordered Nonlinear Systems}
\author{Joshua D. Bodyfelt$^{1,2,3}$, Tsampikos Kottos$^{1}$ and Boris Shapiro$^4$}

\affiliation{$^1$Department of Physics, Wesleyan University, Middletown, Connecticut 06459, USA \\
$^2$MPI for Dynamics and Self-Organization, Bunsenstr.\ 10, 37073 G\"ottingen, Germany\\
$^3$MPI for the Physics of Complex Systems, N\"othnitzer Stra\ss e 38, D-01187 Dresden, Germany\\
$^4$Technion - Israel Institute of Technology, Technion City, Haifa 32000, Israel}

\begin{abstract}
We show, using detailed numerical analysis and theoretical arguments, that the normalized participation 
number of the stationary solutions of disordered nonlinear lattices obeys a one-parameter scaling law. 
Our approach opens a new way to investigate the interplay of Anderson localization and nonlinearity based 
on the powerful ideas of scaling theory.
\end{abstract}


\maketitle

%%%%%%%%%%%%%%%%%%%%%%%%%%%%     Introduction     %%%%%%%%%%%%%%%%%%%%%%%%%%%%%%%%%%%%%
{\it Introduction:} Wave propagation in naturally occurring or engineered complex media is an interdisciplinary
field of research that addresses systems as diverse as classical, quantum and atomic-matter waves. Despite this
diversity, the wave nature of these systems provides a common framework for understanding their transport
properties and often leads to new applications. One such characteristic is wave interference phenomena. Their
existence results in a complete halt of wave propagation in random media, which can be achieved by increasing the
randomness of the medium. This phenomenon was predicted fifty years ago in the framework of quantum (electronic)
waves by Anderson \cite{A58} and its existence has been confirmed in recent years in experiments with classical
\cite{WBLR97,CSG00,SGAM06,BZKKS09,HSPST08,C99,LAPSMCS08,PPKSBNTL04,SBFS07} and matter waves \cite{A08,Ignuscio}.

In many of these experiments, the appearance of nonlinearities,
induced either due to the nonlinear Kerr-effect (in the framework
of nonlinear wave propagation in disordered photonic lattices)
\cite{SBFS07,LAPSMCS08,PPKSBNTL04} or due to atom-atom interactions 
(in atomic transport of Bose-Einstein Condensates in optical lattices) 
\cite{A08, Ignuscio}, might affect drastically
the phenomenon of Anderson localization. An important question is
therefore how the interplay between disorder and nonlinearity
might complement, frustrate, or reinforce each other
\cite{B89,A92,B95,MT95}. This notion is not only of experimental
importance, but it raises a number of unsolved theoretical
questions as well. The theoretical study of localization in random
nonlinear lattices has been advanced using several approaches; including
the studies of transmission\cite{GK92}, wavepacket dynamics
\cite{PS08}, and stationary solutions
\cite{FSW86,KA00,FIM08}.

In this Letter, we approach the interplay of nonlinearity with
disorder from a different perspective, namely we develop a scaling
theory for localization phenomena in nonlinear random media
described by the Discrete Nonlinear Schr\"odinger Equation
(DNLSE). Scaling ideas played a major role in understanding
various properties of {\it linear} disordered systems, including
the structure of their eigenstates \cite{EM08,CMI90,FM92,CGIFM92}.  
However, solutions of the DNLSE, for sufficiently strong 
nonlinearity, have nothing in common with the solutions of the 
linear problem. Indeed, the number of solutions of the DNLSE 
is generally much larger than the number of eigenstates of the
corresponding linear problem and, in particular, many solutions 
appear outside the spectrum of the linear system. It is therefore 
quite remarkable that the scaling ideas can be extended to the 
nonlinear case. Specifically, we find that the rescaled participation 
number $p_N(\chi)$ of the stationary solutions of the DNLSE of lattice 
size $N$ and nonlinearity $\chi$, obeys a one-parameter scaling, i.e.
\begin{equation}
{\partial p_N(\chi)\over \partial \ln N} = \beta \left(p_N(\chi)\right)\quad {\rm where} \quad p_N(\chi)=
{\langle \xi_N(\chi)\rangle \over \langle \xi_N^{\rm ref}(\chi)\rangle}
\label{eq1}
\end{equation}
Above $\beta$ is a {\it universal} function of $p_N$ alone, which is independent of any microscopic properties
of the system under investigation, and $\langle \cdots\rangle$ denotes an averaging over disorder realizations and
over states within a small frequency window. The participation number $\xi_N$ is \cite{KM93}
\begin{equation}
\label{eq4}
\xi_N\equiv {1\over \sum_{n=1}^N|\psi_n|^4}
\end{equation}
and is proportional to the effective number of nonzero components
$\psi_n$ of a stationary solution of the DNLSE. $\xi_N^{\rm
ref}$ is the participation number for some reference ensemble,
chosen such that it supports the most extended states for
a specific lattice topology and nonlinearity. It will be argued
below that $\xi_N^{\rm ref}$ is proportional to the system size
$N$. Note, however, that it would not be accurate to replace
$\xi_N^{\rm ref}$ by $N$ because the coefficient of proportionality 
between the two lengths depends weakly on $\chi$.
Eq.(\ref{eq1}) is confirmed in the following via detailed numerical 
simulations with quasi one-dimensional ($1D$) disordered systems described by
Banded Random Matrix (BRM) models and is supported by theoretical
arguments \cite{note1}.

%%%%%%%%%%%%%%%%%%%%%%%%%%%%%%%%%%%%%%%%%%%%%%%%%%%%%%%%%%%%%%%%%%%%%%%%%%%%%%%%%%%%%%%%%%%%%%%%%%%%%%%5
{\it Mathematical Model:} We consider a class of {\it random} systems described by
the time-independent DNLSE:
\begin{equation}
\label{eq2}
\sum_m H_{nm} \psi_m + \chi |\psi_n|^2 \psi_n=\omega \psi_n \,\,,
\end{equation}
where $\omega$ are the frequencies of the stationary solutions and
$\psi_n$ is their amplitude at site $n$. The connectivity matrix
$H$ defines the topology of the sample. In the case
of strictly $1D$ disordered systems, it is a three-diagonal matrix with
$H_{n,n\pm1}=-1$ while $H_{nn}$ are random independent variables
given by some distribution. In our simulations below, we consider
the challenging case where $H$ belongs to a BRM ensemble,
having in mind quasi-$1D$ systems with $b$ propagating modes \cite{CMI90,FM92}.

The BRM ensemble is defined as a set of real symmetric $N\times N$ matrices with elements
$H_{nm}=0$ for $|n-m|\geq b$, while inside the bandwidth $b$ they are independent random variables given by a
Gaussian distribution of mean zero and fixed variance \cite{CMI90}
\begin{equation}
\label{eq3}
\langle H_{nm}\rangle=0,\quad \langle H_{nm}^2 \rangle={(N+1)(1+\delta_{nm})\over b(2N-b+1)}.
\end{equation}
With this normalization, the eigenvalues of $H$ (in the limit of large $N,b$) are located in the
interval $(-2,2)$ \cite{CMI90}.

{\it Numerical Method:} The stationary solutions of the nonlinear Eq.(\ref{eq2}) were found numerically by
utilizing a continuation method approach. Starting with the linear modes of $H$ as an initial guess, we take
a small step in $\chi$ ($\delta \chi\sim 10^{-4}$), and solve the nonlinear system of Eq.(\ref{eq2}). This
is achieved by minimizing a multivariable $(N+1)$-dimensional vector function ${\vec F}_n=\sum_m H_{nm}\psi_m+
\chi|\psi_n|^2\psi_n-\omega\psi_n ,n=1,2,...N$ and ${\vec F}_{N+1}=\sum_n|\psi_n|^2-1$. Using this method we 
find $(\{\psi_n\};\omega)$ with tolerance $10^{-8}$. The resulting solution then becomes the initial guess 
for the nonlinear solver for the next step in $\chi$.

\begin{figure}[htb]
\includegraphics[width=\columnwidth,keepaspectratio,clip]{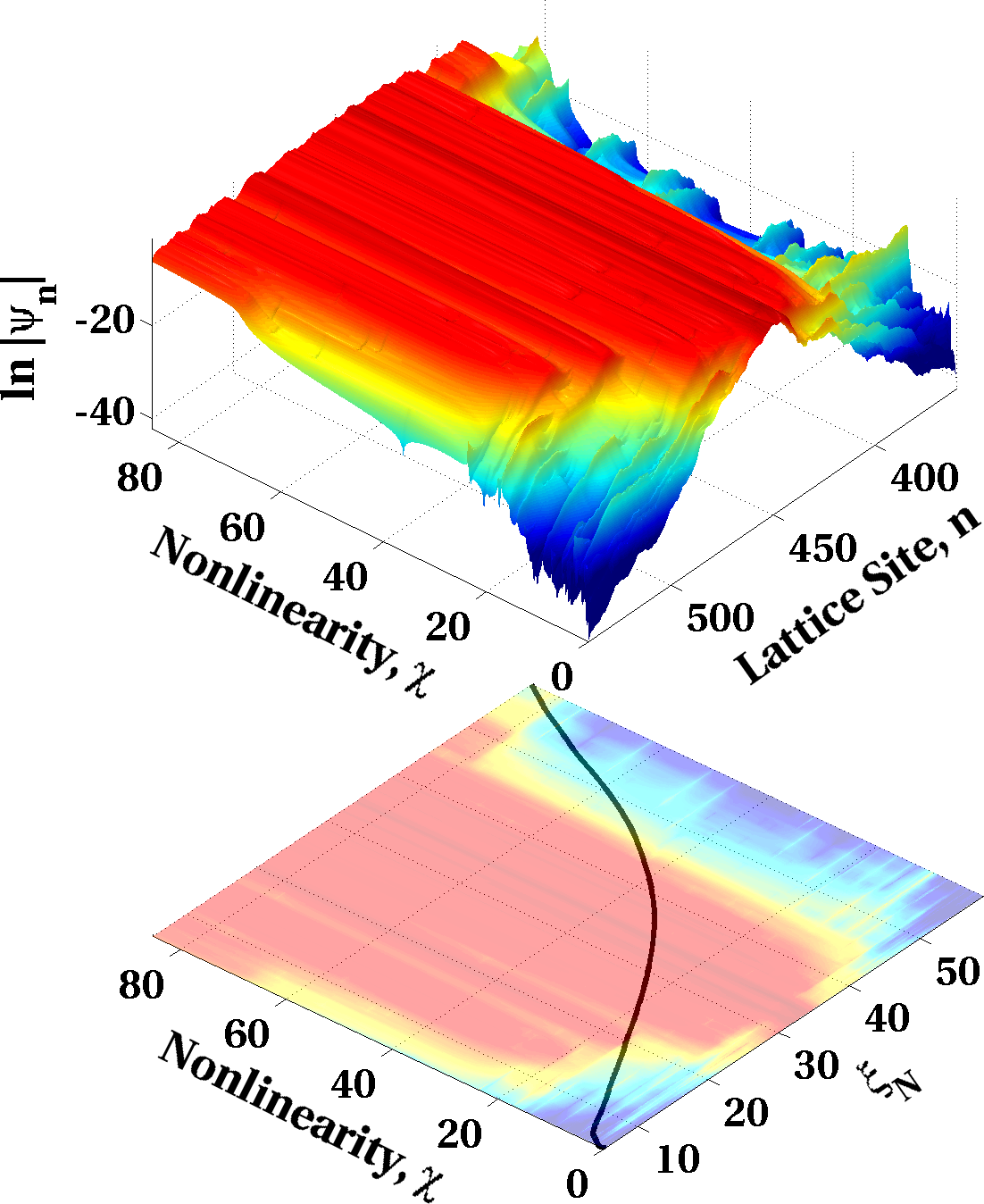}
\caption{
\label{fig1}
Upper panel: Parametric evolution as the nonlinearity $\chi$ increases 
of a stationary solution of the BRM, Eq.(\ref{eq2}), with $b=4$.
For $\chi=0$ the mode is exponentially localized while as $\chi$
increases it becomes delocalized over the whole lattice. Lower
panel: The participation number $\xi_N(\chi)$ (solid black line) 
as a function of $\chi$. The coloring reflects the wavefunction intensity 
shown in the upper panel. }
\end{figure}

An "evolution" of a representative Anderson localized mode as nonlinearity $\chi$ increases is reported in Fig.~\ref{fig1}.
 We observe that while initially (small $\chi$ values) its shape remains unaffected, eventually
it starts delocalize until it spreads over the whole sample. The
degree of delocalization as a function of $\chi$ is reflected in
the behavior of the participation number $\xi_N(\chi)$ (lower
panel of Fig.~\ref{fig1}). We want to investigate how the average
participation number $\langle \xi_N\rangle$ of the stationary
modes of Eq.(\ref{eq2}) is affected by nonlinearity. The
averaging $\langle\cdots\rangle$, has been performed over
solutions with corresponding $\omega$ being inside a small
frequency interval (below $\omega\in [-1,1]$) such that the nature
of wavefunctions is statistically the same. For better statistical
processing, a number of disorder realizations has been used, such
that the total number of obtained stationary solutions is at least
$10^4$.

%%%%%%%%%%%%%%%%%%%%%%%%%%%%%%%%%%%%%%%%%%%%%%%%%%%%%%%%
{\it Nonlinear Localization Length}: We start our analysis by introducing the asymptotic localization length 
$\lambda_{\chi}$ defined via the participation number $\xi_N$:
\begin{equation}
\label{eq5}
\lambda_{\chi}=\lim_{N\rightarrow \infty} \langle \xi_N(\chi)\rangle.
\end{equation}

\begin{figure}[htb]
\includegraphics[width=\columnwidth,keepaspectratio,clip]{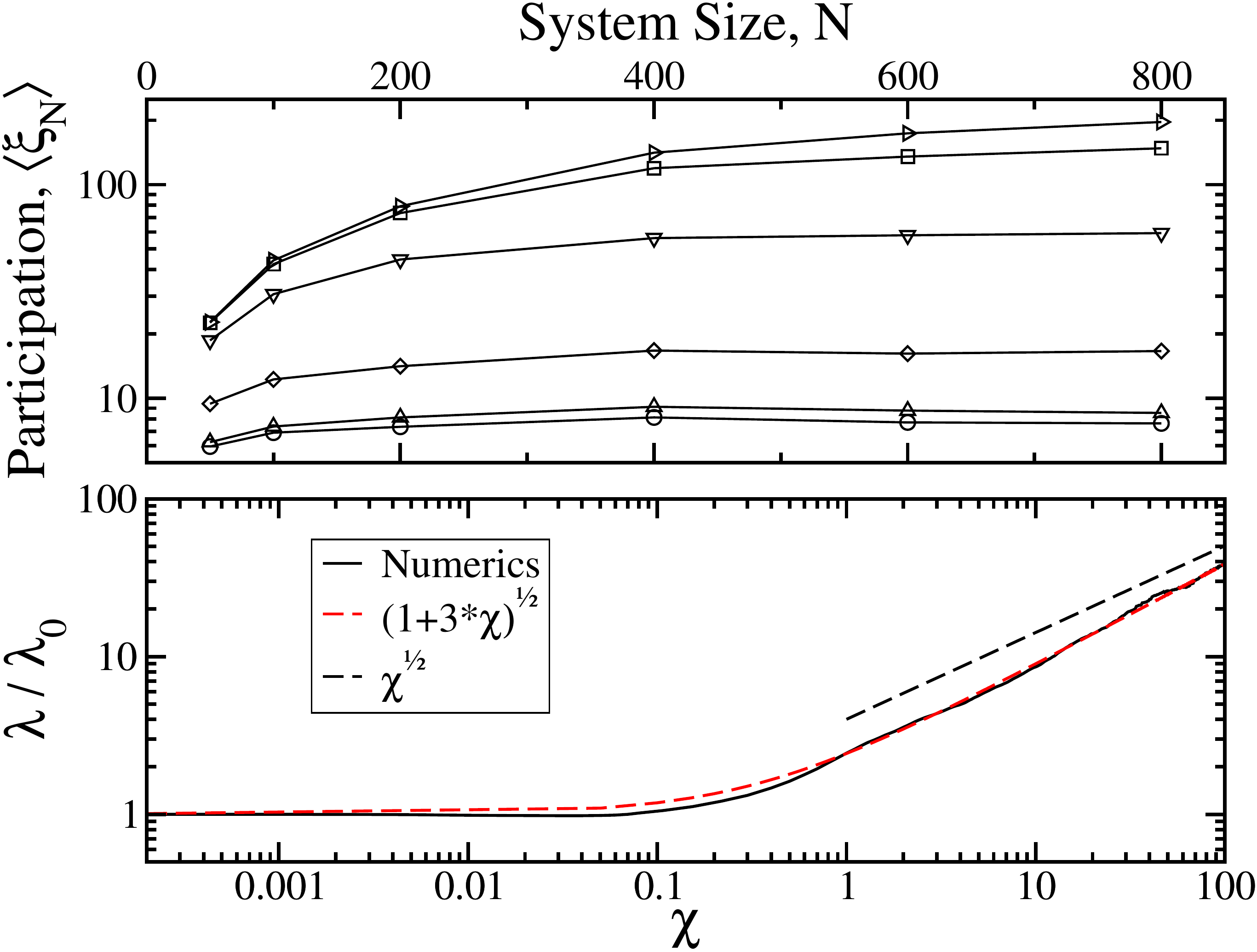}
\caption{
(a) The participation number $\xi_N(\chi)$ vs. the system size $N$, for various nonlinearity strengths: $\chi=0\;(\circ), 
\chi=0.1\;(\bigtriangleup), \chi=1\;(\lozenge), \chi=10\;(\bigtriangledown), \chi=50\;(\square), \chi=100\;(\rhd)$.
The asymptotic localization length $\lambda_{\chi}$ is evaluated by a direct fit of the saturation value of $\xi_N
(\chi)$ in the limit $N\rightarrow \infty$; (b) The extracted $\lambda_{\chi}$ versus the nonlinearity strength
$\chi$. The black dashed line has slope 0.5 and is drawn in order to guide the eye. The red dashed line is the
best fitting curve Eq.(\ref{eq7}).
}\label{fig2}
\end{figure}

In Fig.~\ref{fig2}a we report some representative data for the participation number $\langle\xi_N(\chi)\rangle$, 
as a function of the system size $N$. From these plots we extract the saturation value
$\lambda_{\chi}$. The resulting data are summarized in Fig.~\ref{fig2}b by referring to the rescaled localization
length $\lambda_{\chi}/\lambda_0$, where $\lambda_0$ is the localization length given by Eq.(\ref{eq5}) for the
linear (i.e. $\chi=0$) system. We find that
\begin{equation}
\label{eq6}
{\lambda_{\chi}\over\lambda_0}\propto
\left\{
  \begin{array}{lcr}
1 & {\rm for}           & \chi<\chi^*\\
\sqrt{\chi} & {\rm for} & \chi\gg \chi^*
\end{array}\right.
\end{equation}
where $\chi^*\sim 0.3$ \cite{PS08}. A simple interpolation formula which agrees with our numerical results (see Fig.~\ref{fig2}b) is:
\begin{equation}
\label{eq7}
\lambda_{\chi}=\lambda_0 \sqrt{1+a_0\chi}
\end{equation}
where the fitting parameter $a_0$ was found to be $a_0\approx 3$.
While interpreting the above equations, it is crucial to keep in
mind that all the lengths depend on frequency $\omega$ - when
comparing these lengths for different values of the nonlinearity
$\chi$, one should keep $\omega$ (approximately) fixed.

The following heuristic argument provides some understanding of the behavior of $\lambda_{\chi}$ in the two 
limiting cases. First, let us note that the inverse of the participation number $\xi_N^{-1}(\chi)$ is proportional 
to the interaction energy stored in the system described by the DNLSE, 
i.e. $E_{\rm int}= {\chi\over 2}\sum_n |\psi_n|^4$ . For $\chi=0$ the localization length $\lambda_{\chi}$ 
is equal to $\lambda_0$, while due to normalization the corresponding stationary solutions of Eq.(\ref{eq2}) are 
$\psi_n^{(0)} \sim 1/\sqrt{\lambda_0}$. When $\chi$ increases, $E_{\rm int}$ grows within the localization length $\lambda_0$, 
but $\omega$ is assumed to be approximately fixed. This increase in $E_{\rm int}$ is compensated by further 
spreading of the wavefunction beyond $\lambda_0$. The
first question is: what is the value of the nonlinearity strength
$\chi^*$, for which the spreading beyond $\lambda_0$ becomes
significant? An estimation is achieved by comparing the
interaction energy $E_{\rm
int}=(\chi/2)\sum_n|\psi_n^{(0)}|^4\approx (\chi/2)\lambda_0
(1/\lambda_0^2) = \chi/2\lambda_0$ stored in the "localization
box" of size $\lambda_0$ to the corresponding mean level spacing
$\Delta_{\lambda_0} \sim 1/\lambda_0$. From this comparison, we
get $\chi^*\sim 1$.

Next, we consider the limit $\chi\gg \chi^*$, when $\lambda_{\chi}\gg\lambda_0$. In this case, the wavefunctions
$\psi^{(\chi)}\sim 1\sqrt{\lambda}$ spread over $\lambda_{\chi}/ \lambda_0$ "localization boxes". We
make the following self-consistent argument: The total interaction energy stored is $E_{\rm int}= (\chi/2)
\sum_n|\psi_n^{(\chi)}|^4\approx (\chi/2)\lambda_{\chi} (1/\lambda_{\chi}^2) = \chi/2\lambda_{\chi}$. The
interaction energy per box is therefore $(\chi/2\lambda_{\chi})(\lambda_0/\lambda_{\chi}) = (\chi/2)\lambda_0/
\lambda_{\chi}^2$. However, from the previous considerations we know that one localization box can "resist"
an energy $\Delta_{\lambda_0} \sim 1/\lambda_0$. A self-consistency condition gives $\lambda_{\chi}\sim
\lambda_0 \sqrt{\chi}$, which agrees with the $\chi\gg \chi^*$ asymptotic in Eq.(\ref{eq6}).

{\it Reference Ensemble}: The most ergodic stationary solutions of Eq.(\ref{eq2}) correspond to connectivity
matrices $H$, taken from the GOE \cite{note3}. These solutions will spread over the entire system, of size
$N$. However, this does not mean $\xi_N^{\rm ref}=N$, since $\psi_n^{(\chi)}$ might have some (oscillatory)
structure. We therefore assume that
\begin{equation}
\label{eq8}
\xi_N^{\rm ref}(\chi) = N \alpha(\chi)\quad {\rm where} \quad   1/3\leq\alpha(\chi)\leq 1.
\end{equation}
The lower value of $\alpha (\chi)$ is achieved in the limit of
$\chi\rightarrow 0$, where $\langle \xi_N^{\rm ref}\rangle =N/3$
\cite{CMI90,note3}. The fact that $\langle \xi_N^{\rm ref}\rangle$
is $3$ times less than the system size is due to Gaussian
fluctuations in the components $\psi_n$. The other limiting case
of $\chi\rightarrow \infty$ corresponds to $\alpha=1$. Indeed,
strong nonlinearity favors a completely uniform $|\psi_n|^2$
(again, for fixed and moderate $\omega$). The following
argument, similar in spirit to the "maximum entropy" ansatz, can
provide some understanding. Let us denote the components
$\psi_n^{(\chi)}$ of a stationary solution by a random variable
$x$. Assume that the variable $x$ follows a distribution ${\cal
P}(x)$, "as random as possible" but with the constraint ${\cal
N}=\int dx x^2 {\cal P}(x) = 1/N$ dictated by normalization of the
wavefunction. For $\chi=0$, it was shown that the "most random"
distribution is the Gaussian. It can be found by
maximizing the entropy $S=-\int dx {\cal P}(x) \ln{\cal P}(x)$
with the above constraint \cite{AL86}. For $\chi\neq 0$, one 
however has also to consider the increase of interaction energy
$E_{\rm int}$. The entropy favors a distribution ${\cal P}(x)$,
"as random as possible" for the values of $x$ at various sites;
e.g. the distribution of the "weight" when all of the probability 
is on a single site is as likely as the configuration with equal values of
all the sites. The energy however favors a uniform distribution -
clearly, a "single site weight" will require huge energy. Thus,
the correct quantity to minimize is the "free energy" functional 
$F[{\cal P}]= -S+ E_{\rm int} + \beta {\cal N}$. We get
\begin{equation}
\label{eq9}
{\cal P}(x)=C \exp(-\beta x^2 - (1/2) \chi x^4)
\end{equation}
where $C$ and $\beta$ are defined from the constraints ${\cal
N}=1/N$ and $\int {\cal P}(x) dx =1$. For $\chi=0$ the
Gaussian distribution is recovered, whereas for
$\chi\rightarrow \infty$ the distribution turns $\delta$-like
around $x=1/\sqrt{N}$.

%%%%%%%%%%%%%%%%%%%%%%%%%%%%%%%%%%%%%%%%%%%%%%%%%%%%%%%%
\begin{figure}[htb]
\includegraphics[width=\columnwidth,keepaspectratio,clip]{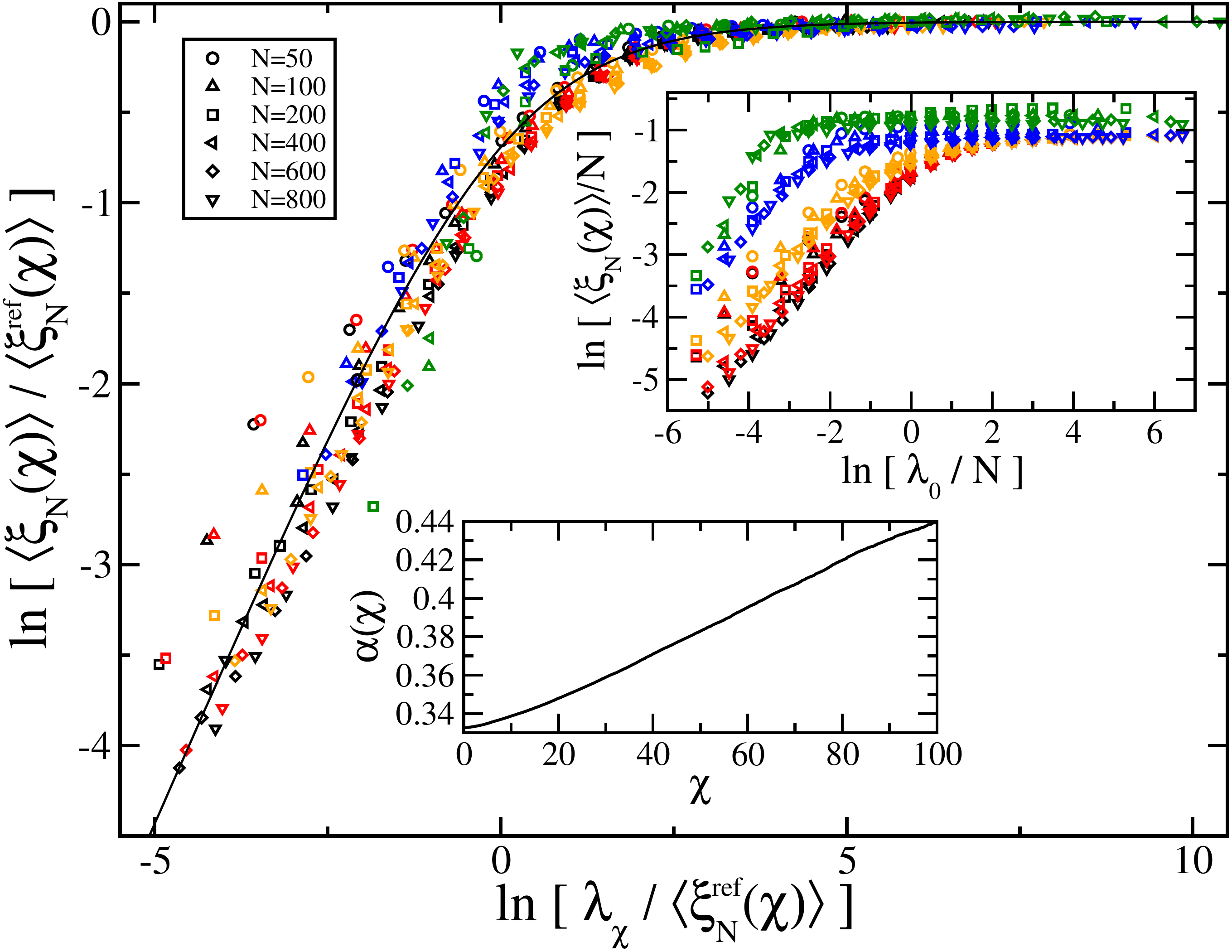}
\caption{Main panel: Participation ratio $\xi_N(\chi)/\xi_N^{\rm ref}(\chi)$ for the BRM model vs. the rescaled
parameter $\lambda_{\chi}/\xi_N^{\rm ref}(\chi)$ for various $(N,b)$-values (for each $N$ we have used at least 
$12$ different $b$ values in the interval $1\leq b \leq N/2$). Colors correspond to nonlinear values
of: $\chi=0$ (black), $\chi=0.1$ (red), $\chi=1$ (orange), $\chi = 10$ (blue), and $\chi = 100$ (green).
The solid line is the best fit with the function $y=0.98x/(1+0.98x)$. Upper inset: The same data as in the main
panel, reported using the scaling variables $\langle\xi_N\rangle/N$ and $\lambda_0/N$. Lower Inset: The numerically
evaluated $\alpha(\chi)$ vs. $\chi$ for the nonlinear GOE reference ensemble. The size of the matrix $H$ is 
$N=800$. In the limit $N\rightarrow \infty$ and for $\chi\gg 1$, the function $\alpha(\chi)\rightarrow 1$.
}
\label{fig3}
\end{figure}

%%%%%%%%%%%%%%%%%%%%%%%%%%%%%%%%%%%%%%%%%%%%%%%%%%%%%%%%%%%%%%%%%%%%%%

{\it One Parameter Scaling Ansatz}: We are now equipped to formulate a scaling theory for the stationary solutions
of Eq.(\ref{eq2}). The scaling ansatz of Eq.(\ref{eq1}) is equivalent to postulating the existence of a function
$f(x)$ such that
\begin{equation}
\label{eq10}
{\langle \xi_N(\chi)\rangle \over \langle\xi_N^{\rm ref}(\chi)\rangle}=f(x)\quad {\rm where} \quad
x={\lambda_{\chi}\over \langle\xi_N^{\rm ref}(\chi)\rangle}
\end{equation}
In the delocalized limit $x\gg 1$, the function $f(x)$ approaches the saturation value $1$. On the other hand,
when $N\rightarrow \infty$ (i.e. $x\ll 1$) we have that $\langle \xi_N(\chi)\rangle \rightarrow \lambda_{\chi}$, thus $f(x)
=x$. We have numerically tested Eq.(\ref{eq10}), for the DNLSE using a BRM ensemble for the
connectivity matrix $H$. Various values of $N$ and $b$ in the ranges $50\leq N\leq 800$ and $1\leq b\leq N/2$ were
used in the analysis. The numerical data are reported in Fig.~\ref{fig3}, confirming nicely the scaling
ansatz of Eq.(\ref{eq10}). Finally, it is reassuring that Eq.~(\ref{eq10}) in the limit $\chi=0$, recovers the 
scaling relation found for linear disordered lattices \cite{CMI90,FM92,CGIFM92}.

It is clear that "strong" Discrete Breathers (DB) localized at few sites are excluded from our above considerations. 
In fact, such solutions correspond to large frequency values $\omega\sim \chi$, which for very large 
$\chi$ are always outside {\it any} fixed small frequency window over which our scaling analysis is performed. 
The analysis of DB can be done separately, since they originate from the clean system with large $\chi$ and 
$\omega$, in which disorder has only a minor effect. Thus, all lengths appearing in Eq.(\ref{eq10}) are (by 
their definition) approximately the same, i.e.  $\xi_N^{\rm ref}\sim \xi_N(\chi) \sim\lambda_{\chi}$, on 
the order of the lattice spacing. This is a rather trivial limit, where Eq.(\ref{eq10}) is again expected to be
approximately valid. 

{\it Conclusions}: We presented a one-parameter scaling theory for the average participation number of the 
stationary solutions of low-dimensional disordered nonlinear systems described by the DNLSE. Via numerical 
analysis and theoretical considerations, we have established Eq.(\ref{eq1}) which allows us to conclude that 
changing disorder strength, nonlinearity, and frequency, in the way described by Eq.(\ref{eq10}), would not 
change the (average) spatial extent of the stationary states of the DNLSE. The one-parameter scaling theory 
presented here is a powerful approach in the quest of understanding the interplay between nonlinearity and 
disorder. Although the focus of this Letter was on the structure of stationary modes, our approach can further 
be used to understand various other observables \cite{BKS09}.

%-----------------------------------------------------------------------------------------------
This research was supported by a grant from the United States-Israel Binational Science
Foundation (BSF), Jerusalem, Israel, and by the DFG FOR760 "Scattering Systems with Complex Dynamics".
%%%%%%%%%%%%%%%%%%%%%%%%%%%%%%%%%%%%%%%%%%%%%%%%%%%%%%%%%%%%%%%%%

%%%%%%%%%%%%%%%%%%%%%%%%%%%%%%%%%%%%%%%%%%%%%%%%%%%%%%%%%%%%%%%%%
%%%%%%%%%%%%%%%%%%%%%%%%%%%%%%%%%%%%%%%%%%%%%%%%%%%%%%%%%%%%%%%%%


\begin{thebibliography}{99}
\bibitem{note1}However, all our results can be immediately translated to the case of strictly $1D$ disordered systems.
\bibitem{note3}For strictly $1D$ tight-binding models the reference system has a three-diagonal connectivity
matrix $H$, with $H_{nn}=0$. In this case, for $\chi=0$ we get $\xi_N^{\rm ref}=2N/3$ \cite{CGIFM92}.
\end{thebibliography}
\end{document}